\documentclass{aa}

\usepackage{natbib,twoopt}
\usepackage[breaklinks=true, backref=page]{hyperref} 
\bibpunct{(}{)}{;}{a}{}{,}             
\makeatletter
  \newcommandtwoopt{\citepads}[3][][]{\href{http://adsabs.harvard.edu/abs/#3}%
    {\def\hyper@linkstart##1##2{}%
     \let\hyper@linked\@empty\citepalp[#1][#2]{#3}}}
\makeatother

\usepackage{graphicx}
\usepackage{txfonts}
\usepackage{esint}
\usepackage[capitalise]{cleveref}
\usepackage{booktabs}
\usepackage{multirow}
\usepackage{float}
\usepackage{dsfont}

\usepackage{tikz}
\usetikzlibrary{arrows.meta}
\usetikzlibrary{shapes.arrows}
\usetikzlibrary{calc}
\usepackage{siunitx}
\usepackage{pgfplots}
\pgfplotsset{compat=1.17}
\usepackage{ulem}

\renewcommand*{\d}[1]{\operatorname{d}\!{#1}}

\makeatletter
\renewcommand*\aa@pageof{, page \thepage{} of \pageref*{LastPage}}
\makeatother

\pgfmathdeclarefunction{gauss}{2}{%
    \pgfmathparse{1/(#2*sqrt(2*pi))*exp(-((x-#1)^2)/(2*#2^2))}%
    }

\begin{document} 

\title{Neural posterior estimation for exoplanetary atmospheric retrieval}


\author{%
    Malavika Vasist\inst{1,2}\fnmsep\thanks{F.R.S.-FNRS PhD Research Fellow}\and 
    François Rozet\inst{2}\fnmsep\thanks{F.R.S.-FNRS PhD Research Fellow}\and
    Olivier Absil\inst{1}\fnmsep\thanks{F.R.S.-FNRS Senior Research Associate}\and
    Paul Mollière\inst{3}\and
    Evert Nasedkin\inst{3}\and
    Gilles Louppe\inst{2}
}

\institute{%
    STAR Institute, University of Liège, 19C Allée du Six Août, 4000 Liège, Belgium\\\email{mv.vasist@uliege.be} 
    \and Montefiore Institute, University of Liège, 10 Allée de la Découverte, 4000 Liège, Belgium
    \and Max-Planck-Institut für Astronomie, Königstuhl 17, 69117 Heidelberg, Germany 
}

\date{Received XX July 2022 / Accepted XXX}


\abstract
 {
Retrieving the physical parameters from spectroscopic observations of exoplanets is key to understanding their atmospheric properties.
Exoplanetary atmospheric retrievals are usually based on approximate Bayesian inference and rely on sampling-based approaches to compute parameter posterior distributions. 
Accurate or repeated retrievals, however, can result in very long computation times due to the sequential nature of sampling-based algorithms.}
{We aim to amortize exoplanetary atmospheric retrieval using neural posterior estimation (NPE), a simulation-based inference algorithm based on variational inference and normalizing flows. 
In this way, we aim (i) to strongly reduce inference time, (ii) to scale inference to complex simulation models with many nuisance parameters or intractable likelihood functions, and (iii) to enable the statistical validation of the inference results.
%
}
{We evaluate NPE on a radiative transfer model for exoplanet spectra (\texttt{petitRADTRANS}), including the effects of scattering and clouds. 
We train a neural autoregressive flow to quickly estimate posteriors and compare against retrievals computed with \texttt{MultiNest}.}
{NPE produces accurate posterior approximations while reducing inference time down to a few seconds.
We demonstrate the computational faithfulness of our posterior approximations using inference diagnostics including posterior predictive checks and coverage, taking advantage of the quasi-instantaneous inference time of NPE. Our analysis confirms the reliability of the approximate posteriors produced by NPE.}
{The inference results produced by NPE appear to be accurate and reliable, establishing this algorithm as a promising approach for atmospheric retrieval. Its main benefits come from the amortization of posterior inference: once trained, inference does not require on-the-fly simulations and can be repeated several times for many observations at very low computational cost. This enables efficient, scalable, and testable atmospheric retrieval. }


\maketitle


\section{Introduction}


The characterization of exoplanet atmospheres is concerned with the identification of model parameters that best describe observed exoplanet spectra.
More specifically, atmospheric retrieval aims to relate exoplanet spectra to the parameters of detailed forward models of atmospheric physico-chemical processes \citep{Madhusudhan2018}.
In this setting, Bayesian inference provides a principled framework to identify parameters matching observations. 
The most widely used inference methods for exoplanet retrieval are Markov Chain Monte Carlo (MCMC) algorithms \citep{burningham2017retrieval, madhusudhan2011high, madhusudhan2014h2o, line2013systematic, line2014systematic, wakeford2017hat, evans2017ultrahot, blecic2016observations, ballard2012pursuit} and variants of nested sampling \citep{lavie2017helios, Molli_re_2020, todorov2016water, benneke2013distinguish, waldmann2015rex, waldmann2015tau, oreshenko2017retrieval, macdonald2017hd, gandhi2018retrieval}.
Although both families of sampling-based algorithms are asymptotically exact, their sequential nature is often an obstacle to fast, scalable, and testable inference \citep{cole2021fast}.
First, sampling-based algorithms can take a few hours up to a few days of computation for each single retrieval. Processing just a few observations can quickly add up to several weeks of computing time, which prevents detailed retrievals for large catalogues of observations. With the advent of JWST and of future missions expected to produce a vast number of observations, this becomes largely inapplicable. 
Second, the necessary computational requirements to maintain accurate results often scale poorly with the number of model parameters. 
This issue is especially salient for simulation models that include many nuisance parameters, whose posteriors are typically not of direct interest but need to be computed anyway because sampling-based approaches require sampling the full joint posterior. 
Third, the reliability and statistical rigor of the approximations produced by sampling-based algorithms are difficult to assess. Stastistical validation based on repeated inferences, such as simulation-based calibration \citep{talts2018validating} or expected coverage \citep{hermans2021averting}, are not feasible in a reasonable time.


MCMC and nested sampling inference algorithms also pose fundamental limits to the class of possible simulation models describing the physics of exoplanet atmospheres. 
To operate, they require an explicit and tractable expression of the likelihood function, which generally constrains simulation models to forward processes that are mainly deterministic, or involve only a few nuisance parameters.
Yet, when data quality will make it possible to account for more details in the cloud physics, more sophisticated simulation models are bound to involve a large number of interfering stochastic processes, resulting in an implicit or intractable likelihood. Possible examples include the cloud formation mechanisms \citep[e.g., via seeding by nucleation,][]{Lee_2018}, their growth \citep[e.g., via coagulation/surface growth,][]{helling2006dust}, their diffusion processes and interactions with the surrounding thermodynamic conditions \citep[e.g., by settling and mixing,][]{woitke2020dust}, or their evolution with time (e.g., by ionisation).
Retrieval with MCMC or nested sampling becomes impossible in these scenarios, at least not without simplifying assumptions. 

A possible way to speed up the inference process, and thereby allow the introduction of more complex simulation models in atmospheric retrievals, is to rely on recent advances in the field of machine learning. For instance, training a random forest \citep{marquez2018supervised,nixon2022}, a generative adversarial network \citep{zingales2018exogan}, a Bayesian neural network \citep{cobb2019ensemble}, or a convolutional neural network \citep{martinez2022convolutional} to retrieve model parameters from noisy data results in quasi-instantaneous retrieval, after paying the upfront cost of generating a training dataset and of training the network. This comes however at the expense of posterior accuracy, as the resulting parameter distributions are generally not true posteriors in the Bayesian sense, and are sometimes even enforced to follow a multivariate Gaussian distribution. Another approach is to use machine learning to generate more informed, narrower priors \citep{hayes2020optimizing}, or even to replace the exoplanet atmosphere simulator by a surrogate model \citep{himes2022accurate}. These methods have the potential of providing more accurate posterior distributions, but at the expense of a more modest improvement in terms of inference time.

The rapidly developing field of simulation-based inference is now offering new tools to tackle these challenges \citep{doi:10.1073/pnas.1912789117}. For instance, \cite{yip2022sample} perform retrieval using the approach of variational inference with a predefined likelihood, to estimate the posterior distribution of a single spectrum, in an un-amortized fashion. Here, we propose to make use of neural posterior estimation \citep[NPE,][]{papamakarios2016fast, lueckmann2017flexible, greenberg2019automatic}, an approach based on simulation-based inference that makes use of deep learning to amortize the retrieval procedure and bypass the evaluation of the likelihood function. 
With NPE, a neural network learns a surrogate posterior as an observation-parameterized conditional probability distribution, from precomputed simulations over the full prior space of interest.
In this way, retrievals become fast, scalable, and testable.
The rest of the paper is structured as follows.
In Sect.~\ref{sec:methods}, we formalize atmospheric retrieval as a Bayesian inference problem, and describe the NPE approach for approximate inference. We also describe the atmospheric radiative transfer model used in this work. Then, in Sect.~\ref{sec:training}, we describe the setup of our experimental study, present our inference results, compare them against those obtained with nested sampling, and demonstrate their validity using inference diagnostics.
In Sect.~\ref{sec:related} and \ref{sec:conclusion}, we discuss related work and finally conclude our study.


\section{Methods} \label{sec:methods}

\subsection{Simulation-based inference} \label{sec:SBI}

In all generality, simulators are forward stochastic models or computer programs that generate synthetic observations according to input parameters. Formally, a stochastic model takes a vector of parameters of interest $\theta$ as input, samples internally a series of nuisance parameters or latent variables $z \sim p(z | \theta)$ and, finally, produces an observation $x \sim p(x | z, \theta)$ as output, thereby defining an implicit likelihood $p(x | \theta)$. The likelihood is often intractable as it corresponds to
\begin{equation}
    p(x | \theta) = \int p(x, z | \theta) \d{z} = \int p(x | z, \theta) p(z | \theta) \d{z} \, ,
\end{equation}
the integral of the joint likelihood $p(x, z | \theta)$ over the latent space. Even if the likelihood is tractable, which is sometimes the case with physical simulators, the posterior
\begin{equation}
    p(\theta | x) = \frac{p(x | \theta) p(\theta)}{p(x)} = \frac{p(x | \theta) p(\theta)}{\int p(x | \theta') p(\theta') \d\theta'}
\end{equation}
involves an intractable integral over the parameter space, which leads to challenging Bayesian inference problems for simulators with medium to high-dimensional parameter spaces.

These computational obstacles can be bypassed using modern simulation-based inference algorithms. Instead of relying on the likelihood function to perform inference, simulation-based approaches use deep neural networks to parameterize universal density estimators and estimate the posterior.
Among the simulation-based inference algorithms, neural posterior estimation consists in training a conditional normalizing flow $p_\phi(\theta | x)$ with parameters $\phi$ to approximate the posterior distribution $p(\theta | x)$. 
A normalizing flow \citep[][see Appendix \ref{normflow} for further details]{papamakarios2021normalizing} is a composition of invertible and differentiable transformations applied to a simple distribution (e.g., a standard Normal), thereby defining a complex distribution that can be efficiently evaluated and sampled from. 
The transformations are  parametrized by invertible neural networks, making normalizing flows universally expressive parametric distributions that can be trained to approximate other distributions. 
In our case, training is based on amortized variational inference and amounts to the minimization of the expected Kullback-Leibler (KL) divergence between $p(\theta | x)$ and $p_\phi(\theta | x)$ \citep{agakov2004algorithm}, that is
\begin{align}
    \phi^* & = \arg\min_\phi \mathbb{E}_{p(x)} \left[ \text{KL} \big( p(\theta|x) \parallel p_\phi(\theta | x) \big) \right] \nonumber \\
    & = \arg\min_\phi \mathbb{E}_{p(x)} \, \mathbb{E}_{p(\theta | x)} \left[ \log \frac{p(\theta | x)}{p_\phi(\theta | x)} \right] \nonumber \\
    & = \arg\min_\phi \mathbb{E}_{p(\theta, x)} \big[ - \log p_\phi(\theta | x) \big].\label{eq:loss}
\end{align}
Remarkably, the amortization over $p(x)$ of the variational inference objective makes it possible to bypass the sampling or the evaluation of the unknown posterior $p(\theta|x)$ in the second line above. Indeed, the double expectation $\mathbb{E}_{p(x)} \mathbb{E}_{p(\theta | x)}$ can be rewritten as an expectation $\mathbb{E}_{p(\theta, x)}$  over the joint distribution, which we can easily sample in the forward direction as $p(\theta, x) = p(\theta)p(x|\theta)$, regardless of whether the likelihood is tractable or not. 
Once the normalizing flow is trained, evaluating and sampling the posterior density $p_\phi(\theta | x)$ becomes as fast as a forward pass through the network. Inference can be repeated any number of times with different observations, without having to regenerate data from the simulation model. 

\subsection{Atmospheric radiative transfer model} \label{sec:simulator}

The atmospheric model used in this study consists of a deterministic atmospheric forward model implemented with \texttt{petitRADTRANS}, together with a noise model accounting for measurement noise. \texttt{petitRADTRANS} \citep{molliere2019petitradtrans} is a radiative transfer model used to generate emission and transmission spectra for exoplanets with cloudy and clear atmospheres including scattering, as described in \citet{Molli_re_2020}. It includes a parameterized temperature structure and cloud properties. We use the disequilibrium chemistry emission model predefined in \texttt{petitRADTRANS} to compute an emission spectrum based on disequilibrium carbon chemistry, equilibrium clouds, and a spline temperature-pressure profile, defined by 16 parameters in total. We walk through these parameterizations briefly in the following paragraphs. 

The temperature structure uses both freely variable and physically motivated parameterizations based on atmospheric altitudes. The optical depth defined as $\tau$ = $\delta P^\alpha$ is parameterized as a function of the pressure $P$ while keeping $\delta$ and $\alpha$ as model parameters. The temperature structure is split into three parts. The mid altitude (photosphere), with an optical depth $\tau > 0.1$, models the temperature according to the Eddington approximation \citep[Eq.~2 of][]{Molli_re_2020} with $T_{\rm int}$ as a model parameter. In the upper altitude with an optical depth $\tau < 0.1$, the structure is computed by a cubic spline interpolation between $T_1$, $T_2$, and $T_3$ considered as model parameters. In low altitudes (troposphere), wherever the atmospheric temperature gradient of the Eddington approximation is greater than the moist adiabatic gradient (i.e, $\nabla_{\rm edd} > \nabla_{\rm ad}$), convection ensues. The $\nabla_{\rm ad}$ is interpolated from a $T$-$P$-[Fe/H]-C/O space of a chemical equilibrium table. Here the metallicity [Fe/H], and the carbon-to-oxygen number ratio C/O, are also model parameters. 

Once the P-T profile is constructed, equilibrium cloud abundances are calculated in the form of their mass fractions, where they are modified from solar abundances based on the model parameters [Fe/H] and C/O. The cloud mass fractions are further scaled with the scaling parameters $\log \tilde{X}_{\rm Fe}$ and $\log \tilde{X}_{\rm MgSiO_3}$, where $\tilde{X}_{i}= {X}_{0}^{i}/{X}_{\rm eq}^{i}$ is the ratio of the cloud mass fraction $X_0^{i}$ at the cloud base (i.e., at pressure $P_{\rm base}$) to the mass fraction ${X}_{\rm eq}^{i}$ predicted at the same location for the cloud species when assuming equilibrium condensation. The cloud mass fraction decays with altitude based on the settling parameter $f_{\rm sed}$:
\begin{equation}
    X(P) = X_0 \left(\frac{P}{P_{\rm base}}\right)^{f_{\rm sed}} \, .
\end{equation}
For $P>P_{\rm base}$ the cloud mass fraction is zero. The other cloud parameters include the vertical eddy diffusion coefficient $K_{zz}$ and the width of the log normal size distribution $\sigma_{g}$ defined in the \citet{ackerman2001precipitating} cloud model, called Cloud Model 1 in \citet{Molli_re_2020}. The chemical abundances for species H$_2$O, CO, CH$_4$, NH$_3$, CO$_2$, H$_2$S, VO, TiO, PH$_3$, Na, and K are interpolated from the chemical equilibrium table calculated with \texttt{easyCHEM} \citep{molliere2017observing} as a function of $T$-$P$-[Fe/H]-C/O. The model parameter $\log P_{\rm quench}$ is used to account for disequilibrium chemistry through atmospheric mixing. For pressures below $P_{\rm quench}$, the mass fractions of CH$_4$, H$_2$O, and CO are held constant at their values at $P= P_{\rm quench}$.
The gas opacities required for the radiative transfer solution are obtained by combining the correlated $k$ (opacity) tables of individual atmospheric absorbers in the resort-rebin fashion \citep[e.g.,][]{molliere2015model,amundsen2017}. The surface gravity ($\log g$) and radius ($R_p$) of the planet are considered as model parameters to calculate the emission flux. The radiative transfer equations are then solved using the Feautrier method \citep{feautrier1964resolution} as in the self-consistent \texttt{petitCODE} \citep{molliere2015model,molliere2017observing}, which also includes isotropic scattering. Following  \citet{Molli_re_2020}, we rebin down the default wavelength spacing $\lambda/\Delta\lambda = 1000$ to a spacing of 400 between 0.95 to 2.45~$\mu$m. This is done by generating the binned correlated-k opacities in \texttt{petitRADTRANS}, and using them instead of the original opacities to generate linearly binned spectra within the same wavelength range, resulting in vectors of $379$ elements. 
We denote as $f(\theta)$ the output spectrum produced by this first simulation stage, where $\theta$ contains all $16$ model parameters. 

To account for measurement noise and make the simulation model similar to instrumental data, we consider a Gaussian noise model with a standard deviation $\sigma$. The spectra $f(\theta)$ generated by \texttt{petitRADTRANS} are randomly perturbed with additive noise $\epsilon \sim \mathcal{N}(0, \sigma^{2})$, where $\epsilon \in \mathbb{R}^{379}$ is a vector of random noise instances in each wavelength bin. Here we assume the same noise variance in each wavelength bin for the sake of simplicity, but more complex noise models (including noise covariance) could be used in our simulator. The final simulator output is given by $x = f(\theta) + \epsilon$.



\section{Atmospheric retrieval} \label{sec:training}

\begin{figure}
\centering
    \includegraphics[width=\columnwidth]{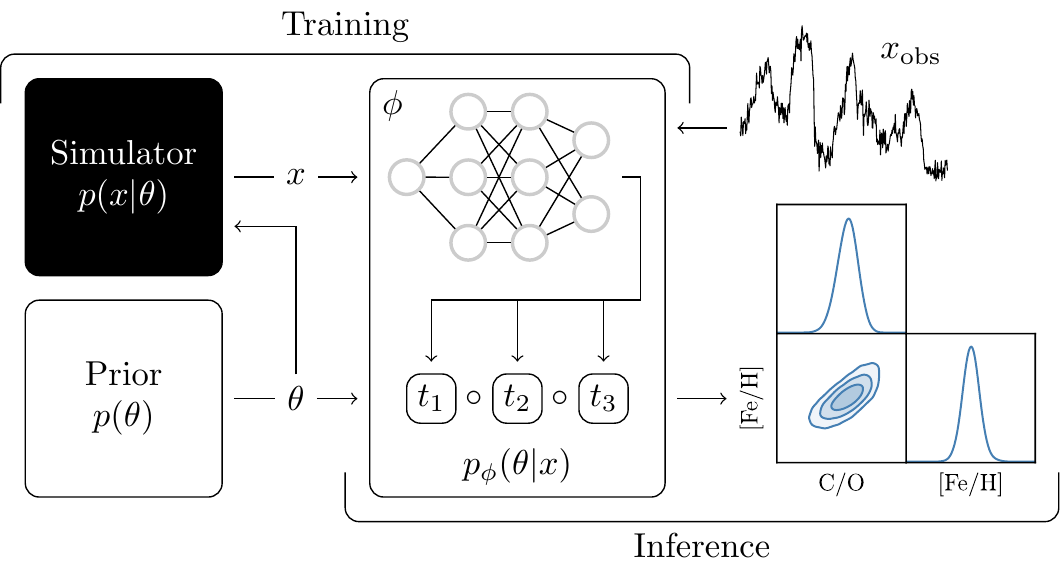}
    \caption{Inference pipeline using amortized neural posterior estimation. The joint simulation model $p(x,\theta)=p(\theta)p(x|\theta)$ is used to generate a training set $\{ (\theta, x) \}$ of model parameters $\theta$ and exoplanet spectra observations $x$. A conditional normalizing flow $p_\phi(\theta|x)$ composed of an embedding network and three invertible transformations $t_i$ is trained to estimate the posterior density $p(\theta|x)$. Once trained, sampling from the posterior estimator is as fast as a forward pass through the normalizing flow. Inference can be repeated for many observations without having to regenerate data nor retrain the normalizing flow.}
    \label{fig:overview}
\end{figure}

We start our empirical evaluation of NPE-based atmospheric retrieval by describing the creation of the training data in Sect.~\ref{sec:exp-setup}, together with the training protocol and a description of the architecture of the neural posterior estimator $p_\phi(\theta | x)$. 
We demonstrate and discuss in Sect.~\ref{sec:exp-retrieval} an example of atmospheric retrieval using NPE, and then validate the statistical quality of the posterior estimation in Sect.~\ref{sec:exp-diagnostics}. Finally, in Sect.~\ref{sec:exp-speedup}, we report and compare computational times against the \texttt{MultiNest} algorithm for nested sampling. 
The inference pipeline is summarized in Fig.~\ref{fig:overview}.

\subsection{Setup}
\label{sec:exp-setup}

\begin{table}[t]
    \caption{Prior distribution over the model parameters.}\label{tab:prior}
    \centering
    \small
    \begin{tabular}{cc|cc}
        \hline \hline
        Parameter & Prior & Parameter & Prior\\
        \hline
        $T_1$ & $\mathcal{U}(0, T_2)$ & $\log \tilde{X}_{\rm Fe}$\tablefootmark{b} & $\mathcal{U}(-2.3,1)$ \\ 
        $T_2$ & $\mathcal{U}(0, T_3)$ & $\log \tilde{X}_{\rm MgSiO_{3}}$\tablefootmark{b} & $\mathcal{U}(-2.3,1)$ \\
        $T_3$ & $\mathcal{U}(0, T_{\rm connect})$ ~\tablefootmark{a} & $f_{\rm sed}$ & $\mathcal{U}(0,10)$ \\
        $\log \delta$ & $P_{\rm phot} \sim \mathcal{U}(10^{-3}, 10^2)$ ~\tablefootmark{c} & $\log K_{zz}$ & $\mathcal{U}(5,13)$ \\
        $\alpha$ & $\mathcal{U}(1,2)$ & $\sigma_{g}$ & $\mathcal{U}(1.05,3)$ \\ 
        $T_{0}$ & $\mathcal{U}(300, 2300)$ $K$ & $R_{P}$ & $\mathcal{U}(0.9,2)$ \\  
        C/O & $\mathcal{U}(0.1, 1.6)$ & $\log g$ & $\mathcal{U}(2,5.5)$ \\  
        Fe/H & $\mathcal{U}(-1.5,1.5)$  & $\log P_{\rm quench}$ & $\mathcal{U}(-6,3)$ \\
        \hline
    \end{tabular}
    \tablefoot{
        \tablefoottext{a}{$T_{\rm connect}$ is the uppermost temperature of the photospheric layer that \texttt{petitRADTRANS} calculates by setting the optical depth 
        $\tau = 0.1$.}
        \tablefoottext{b}{Here $\tilde{X}_{i}= {X}_{0}^{i}/{X}_{\rm eq}^{i}$, where $X_{\rm eq}$ is defined as the mass fraction predicted for the cloud species when assuming equilibrium condensation at the cloud base location.}
        \tablefoottext{c}{$P_{\rm phot}$ is defined as the pressure where the optical depth 
        $\tau = 1$. The parameter $\delta$ is calculated accordingly.}
    }
\end{table}

As a starting point to atmospheric retrieval with neural posterior estimation, we first define in  Table~\ref{tab:prior} a 16-dimensional multivariate uniform prior distribution, with physically motivated ranges for each parameter $\theta$.
This prior distribution is the same as the one used by \citet{Molli_re_2020}. 
Our training data set is composed of $12$ million parameters-spectrum pairs $(\theta, f(\theta))$. It is created by drawing parameters $\theta \sim p(\theta)$ from the prior and passing them through the simulator as shown in Fig.~\ref{fig:overview}. We split this dataset into $90\%$ $9\%$, and $1\%$ for training, validation, and testing respectively.

The posterior estimator $p_\phi(\theta|x)$ is implemented as a neural autoregressive flow \citep{huang2018neural} composed of three transformations. Each transformation is parameterized by a multi-layer perceptron (MLP) with five hidden layers of size \num{512} and ELU activation functions \citep{clevert2015fast}. 
A second network, called the embedding, is used to compress 
the 379-dimensional spectrum $x$ into a vector of 64 features, which is then used to condition the flow with respect to $x$. The rationale behind this compression is that it forces the posterior estimator to extract informative features from the spectra instead of memorizing the training data. The embedding network is implemented as a ResidualMLP (or ResMLP), composed of 10 residual blocks \citep{he2016deep} of decreasing size (two times \num{512}, three times \num{256} and five times \num{128}) and also uses ELU activation functions.
Before training, random noise realizations are added on-the-fly to the spectra to obtain observations $x = f(\theta) + \epsilon$. Following Eq.~\ref{eq:loss}, the flow and embedding networks are trained jointly to minimize the expected negative posterior log-density over the training set. The optimization is carried out through a variant of stochastic gradient descent, namely AdamW \citep{loshchilov2017decoupled}. We use an initial learning rate of $10^{-3}$ that is halved every time the average loss over the validation set has not improved for the last \num{32} epochs, until it reaches $10^{-6}$ to improve training without overfitting \citep{zhang2021dive}. We also use a high weight decay of $10^{-2}$. We train for a total of \num{1024} epochs during which \num{1024} random batches of \num{2048} pairs $(\theta, f(\theta))$ are taken from the training set.

The architectural hyper-parameters were adjusted on validation data. We also explored a neural spline flow \cite{durkan2020contrastive} implementation of the posterior estimator, but in the end implemented a neural autoregressive flow since it gave a lower validation loss. We performed extensive hyper-parameter tuning on the flow and embedding network parameters. For the flow, we explored different numbers of transforms and hidden layer dimensions in the range of $[3,5]$ and $[256,512]$, respectively. For the embedding network, we tried different number of layers in the ResMLP in the range of $[10, 15]$. We also explored different activation functions like ReLU and ELU for both networks.
We explored different values for the initial learning rate and the minimum learning rate in the ranges of $[10^{-5}, 10^{-3}]$ and $[10^{-6}, 10^{-5}]$, respectively. We analyzed the impact of different schedulers like \texttt{ReduceLROnPlateau} and \texttt{CosineAnnealingLR}, available in PyTorch, with patience rates between $[8,32]$. We tried batch sizes between $[2^8, 2^{11}]$ and the number of epochs between $[2^8, 2^{10}]$.
We tuned each hyper-parameter by randomly searching over a grid within their range mentioned above, and studied their impact over $\sim$80 runs in parallel. We selected those that led to lower validation loss and/or more stable training. Amongst all the parameters that we tuned, the parameter weight decay between $[0, 10^{-2}]$ had the most significant impact on the training. We think this is because of the high variance of the input dataset, where some spectra are six orders of magnitude brighter than the rest. This leads to the skewing of the weights to very high values, which is compensated by weight decay to improve training performance. For more details, we refer to the source code of the experiments\footnote{\url{https://gitlab.uliege.be/francois.rozet/sbi-ear}}.

\subsection{Benchmark retrieval} \label{sec:exp-retrieval}

As a demonstration of atmospheric retrieval with NPE, we present inference results for a synthetic exoplanet spectrum $x_\text{obs}$ generated with the parameter values $\theta_\text{obs}$ given in Table~\ref{tab:obs}, similarly to the benchmark retrieval of \citet{Molli_re_2020}. The synthetic spectrum spans a wavelength range from $0.95$ to $2.45$$\mu$m with a continuous wavelength spacing of $\lambda/\Delta\lambda = 400$. As in \citet{Molli_re_2020}, we assume a signal-to-noise ratio of 10 per spectral bin, leading to a standard deviation $\sigma = $0.1257e-17$W m^{-2}\mu m^{-1}$ for the Gaussian noise. The synthetic spectrum used for our retrieval tests is shown in Fig.~\ref{fig:overview}.
    
\begin{table}[t]
    \caption{Parameter values $\theta_\text{obs}$ of the benchmark spectrum $x_\text{obs}$.}\label{tab:obs}
    \centering
    \small
    \begin{tabular}{cc|cc}
        \hline \hline
        Parameter & Value & Parameter & Value \\
        \hline
        $T_1$ & $330.6 K$ & $\log \tilde{X}_{\rm Fe}$ & $-0.86$ \\ 
        $T_2$ & $484.7 K$ & $\log \tilde{X}_{\rm MgSiO_{3}}$ & $-0.65$ \\
        $T_3$ & $687.6$ & $f_{\rm sed}$ & $3$ \\
        $\log \delta$ & $-7.51$ & $\log K_{zz}$ & $8.5$ \\
        $\alpha$ & $1.39$ & $\sigma_{g}$ & $2$ \\
        $T_{0}$ & $1063.6 K$ & $R_{P}$ & $1$ \\
        C/O & $0.55$ & $\log g$ & $3.75$ \\
        Fe/H & $0$ & $\log P_{\rm quench}$ & $-5$ \\
        \hline
    \end{tabular}
\end{table}

\begin{figure*}
   \includegraphics[width=\textwidth]{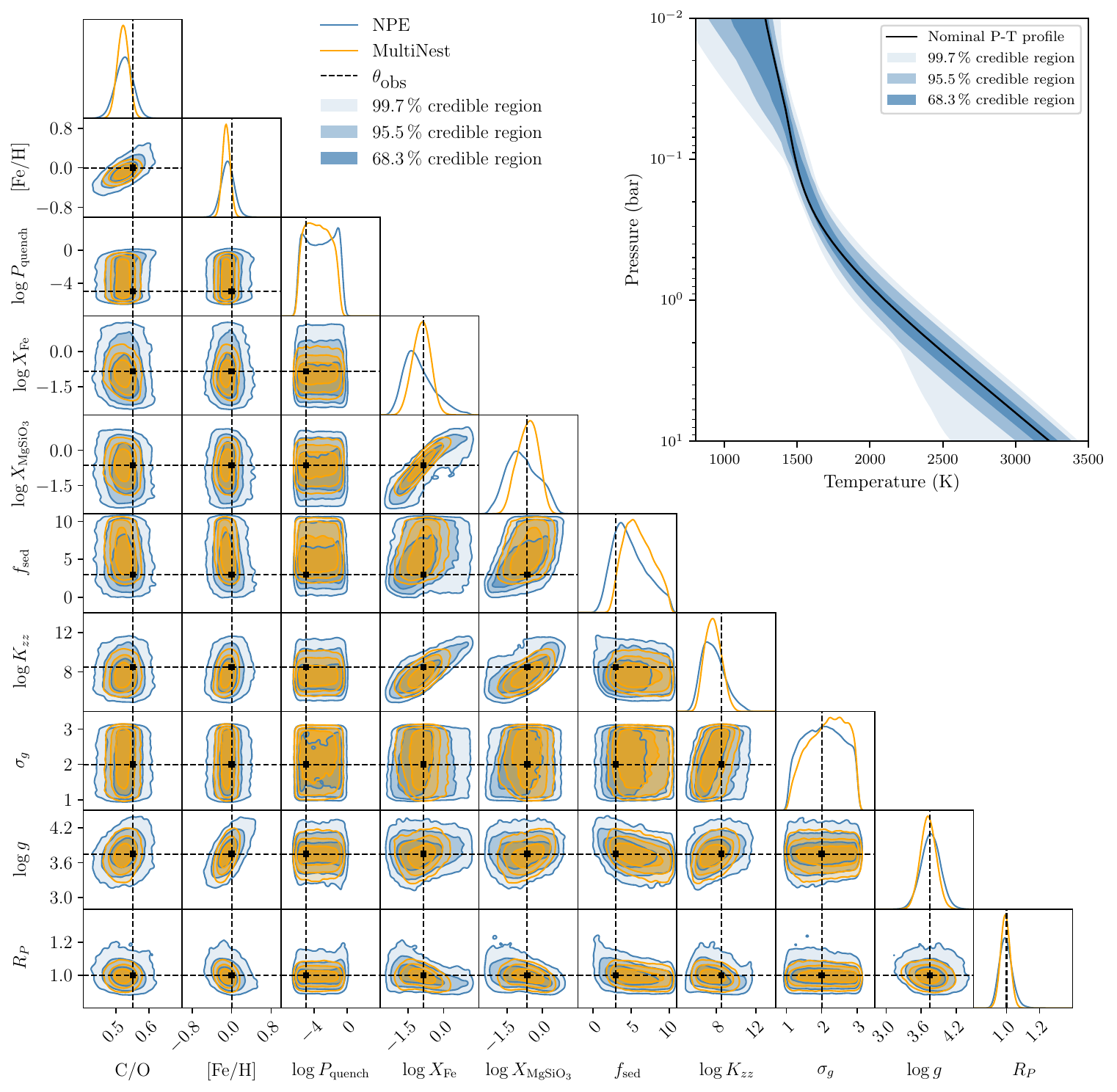}
    \caption{Benchmark retrieval using neural posterior estimation. The corner plot shows 1d and 2d marginal posterior distributions obtained for the benchmark spectrum $x_\text{obs}$ for NPE (in blue) and for nested sampling (in orange). We observe that the nominal parameter values $\theta_\text{obs}$ (in black) are well identified. The top right figure illustrates the posterior distribution of the P-T profiles. }
    \label{fig:results}
\end{figure*}

Retrieval results are summarized in Fig.~\ref{fig:results}. 
The corner plot shows 1d and 2d marginal posterior distributions obtained for the benchmark spectrum $x_\text{obs}$. The marginal posterior distributions are approximated by sampling sufficiently many times the joint posterior distribution from the normalizing flow, which takes only a few seconds to obtain a smooth corner plot.
We observe that the bulk of the marginal posterior distributions (in blue) is centered around the parameter values $\theta_\text{obs}$ (in black) used to generate the spectrum $x_\text{obs}$.
The figure also illustrates the spread in the posterior P-T profiles with respect to the synthetic observation spectrum.
More specifically, we compute and plot posterior P-T profiles for $\theta \sim p_\phi(\theta|x_\text{obs})$ and show their $68.3 \%$, $95.5 \%$ and $99.7 \%$ credible regions.
We see that the P-T profile for $\theta_\text{obs}$ (in black) is constrained mostly within the first credible region of the posterior. These results lead us to believe that the NPE posterior approximation produces coherent posterior distributions. 

\subsection{Validation}
\label{sec:exp-diagnostics}

In Fig.~\ref{fig:results}, we compare the \texttt{NPE} posteriors with those obtained using \texttt{MultiNest} \citep{ferrozhobson2008,ferrozhobson2009,ferrozhobson2013,buchnergeorgakakis2014} for the same noisy synthetic observation, in orange.
While the results obtained with NPE appear to be coherent with respect to the nominal parameter values $\theta_\text{obs}$ and the posterior P-T profiles, we see that the approximate marginal posterior distributions computed with \texttt{MultiNest}, using a sampling efficiency of $0.8$ (recommended for parameter estimation) with 4000 live points, are slightly less dispersed than for NPE. 
On performing several retrievals with different noise realizations (not shown here), it is seen that, each time, the peaks of the individual parameter posterior distributions shift in a similar way in both the retrieval algorithms. This can be seen here in the parameters C/O and Fe/H, similarly shifted slightly to the left. This suggests that these shifts are directly related to the particular noise realization, and that \texttt{MultiNest} and NPE behave in a similar way in presence of noise.

\begin{figure}[t]
    \centering
    \includegraphics[width=\columnwidth]{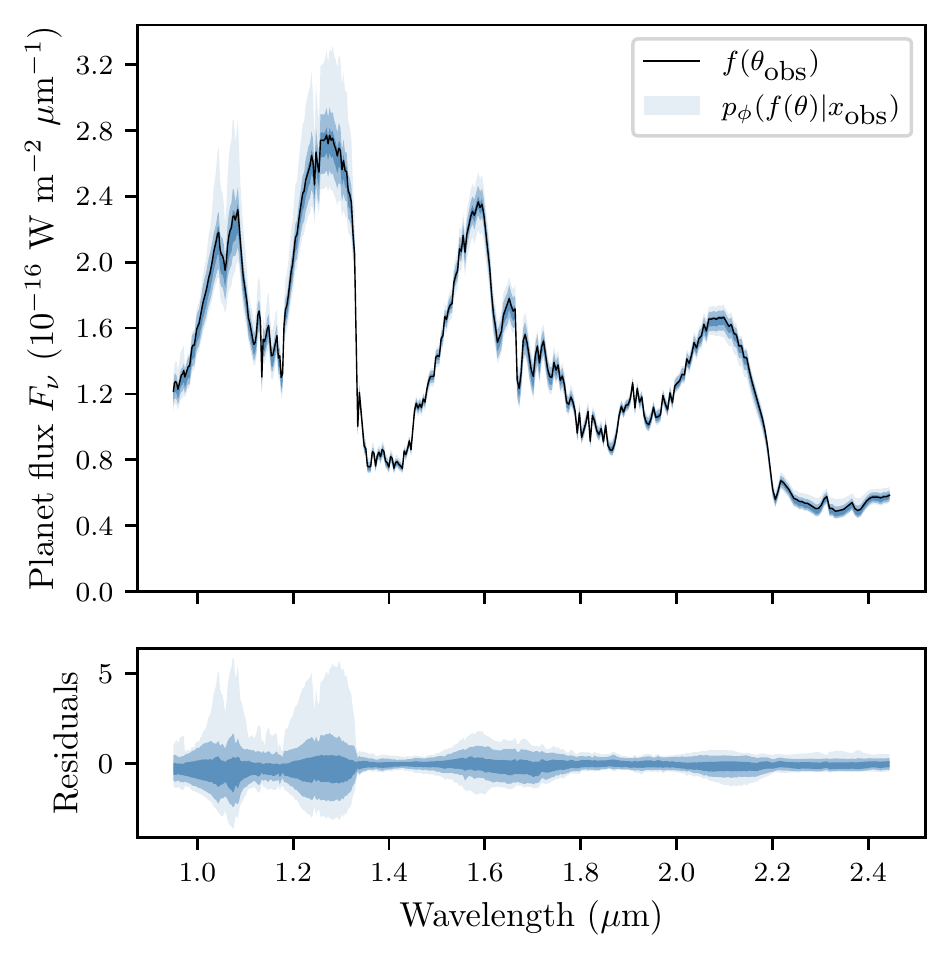}
    \caption{\textit{Top.} Posterior predictive distribution $p(f(\theta)|x_\text{obs})$ of noiseless spectra (without the instrumental noise disturbance $\epsilon$) for the $99.7\%$, $95\%$ and $68.7\%$ quartiles (hues of blue), overlaid on top of the noiseless observed spectrum $f(\theta_\text{obs})$ (black line). {\textit{Bottom.} Residuals of the posterior predictive samples, normalized by the standard deviation of the noise distribution for each spectral channel. 
    }}
    \label{fig:consistency}
\end{figure}

\begin{figure}[t]
    \centering
    \includegraphics[width=\columnwidth]{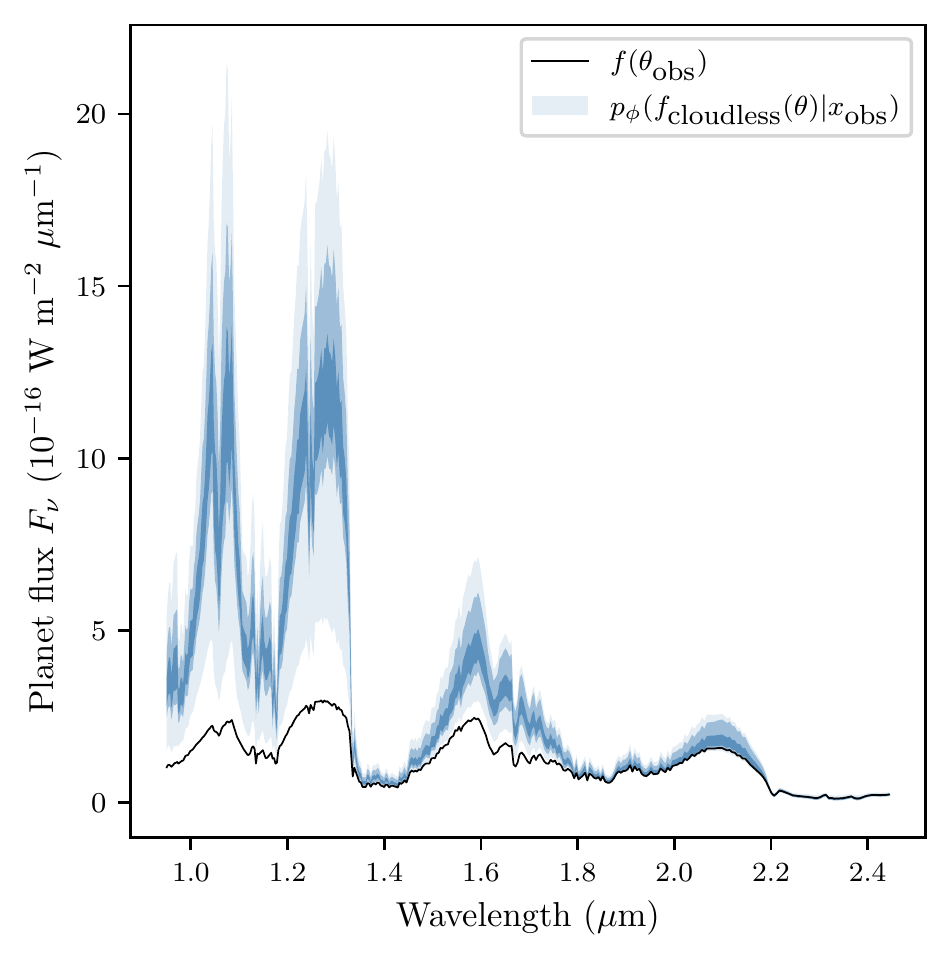}
    \caption{Cloudless realizations of the posterior predictive distribution $p(f_{\rm cloudless}(\theta)|x_\text{obs})$ overlaid on top of $f(\theta_\text{obs})$, where $f_{\rm cloudless}$ artificially sets the cloud scaling factors $\log X_{\rm Fe}$ and $\log X_{\rm MgSiO_{3}}$ to a very small value of $-10$.}  
    \label{fig:consistency_cloudless}
\end{figure}

The difference in the posterior widths for the two algorithms motivates a thorough investigation of the computational faithfulness of the NPE posterior approximations using inference diagnostics, including posterior predictive checks and coverage. We take advantage of the quasi-instantaneous inference of NPE to perform these checks. 
We first perform a quantitative examination of the posterior predictive distribution $p_\phi(f(\theta)|x_\text{obs})$ for spectra without instrumental noise disturbance, which we obtain by sampling parameters from the posterior, $\theta \sim p_\phi(\theta|x_\text{obs})$, and then by computing their spectra $f(\theta)$ with \texttt{petitRADTRANS}. Fig.~\ref{fig:consistency} shows the posterior predictive distribution $p(f(\theta)|x_\text{obs})$ for various quartiles against the noiseless version of the observed spectrum $f(\theta_\text{obs})$.
We observe that (i) the posterior predictive distribution is well constrained, with the 68\% quartile distribution mostly within the $1\sigma$ noise limit as expected, and (ii) that $f(\theta_\text{obs})$ is relatively well centered inside the 68\% quartile along all bins.
Had the posterior distribution $p_\phi(\theta|x_\text{obs})$ been too wide, we would have observed a much larger spread. Had the bulk of the posterior density been at the wrong place, we would have not observed $f(\theta_\text{obs})$ to be well inside the distribution. These reassuring diagnostics are a first indication of the good quality of the inference results obtained with NPE. In particular, they demonstrate that the cloud parameter distributions derived by NPE produce spectra consistent with the observed spectrum. In Fig.~\ref{fig:consistency_cloudless}, we further demonstrate that the parameter values sampled from the somewhat wider NPE cloud posteriors are actually all leading to cloudy solutions, in good agreement with the synthetic input observation. In this figure, we sample parameters from the (cloudy) approximate posterior, but then artificially turn off the clouds, by setting the log of the cloud mass fraction scaling factors $X_{\rm Fe}$ and $X_{\rm MgSiO_{3}}$ to $-10$, to assess their impact on the spectral shape. We see that these cloudless spectra look significantly different from the cloudy ones shown in Fig.~\ref{fig:consistency}. This implies that the posterior predictive distribution samples in Fig.~\ref{fig:consistency} are indeed affected by clouds. 

\begin{figure}[t]
    \centering
    \includegraphics[width=\columnwidth]{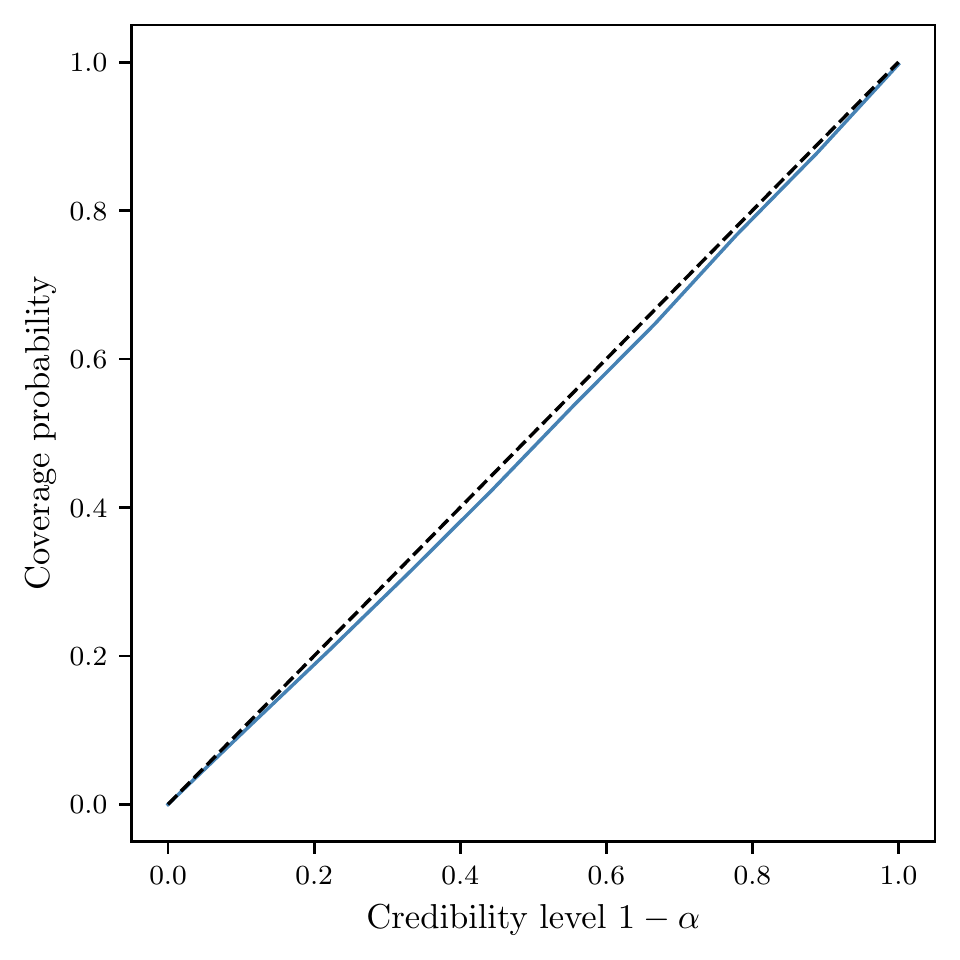}
    \caption{Coverage plot assessing the computational faithfulness of $p_\phi(\theta|x)$ in terms of expected coverage. The coverage probability is close to the credibility level $1-\alpha$, which indicates that the posterior approximations produced by NPE are neither significantly overdispersed (the coverage curve would otherwise be above the diagonal), nor significantly underdispersed (the coverage curve would be below the diagonal).}
    \label{fig:coverage_plot}
\end{figure}

Following \citet{hermans2021averting}, we further evaluate the global computational faithfulness of the NPE posterior approximations in terms of expected coverage. We define the expected coverage probability of the $1-\alpha$ highest posterior density regions derived from the posterior $p_\phi(\theta|x)$ as 
\begin{equation}\label{eq:coverage}
    \mathbb{E}_{p(\theta,x)}\left[\mathds{1}\left(\theta \in \Theta_{p_\phi(\theta|x)}(1 - \alpha)\right)\right],
\end{equation}
where $\mathds{1}(\cdot)$ is the indicator function, and where the function $\Theta_{p_\phi(\theta | x)}(1 - \alpha)$ yields the $1 - \alpha$ highest posterior density region of $p_\phi(\theta | x)$. 
This diagnostic probes the consistency of the posterior estimator $p_\phi(\theta|x)$ and can be used to assess whether the approximate posterior distributions are overdispersed or underdispersed on average.
It is estimated by repeatedly sampling $(\theta,x)$ from the prior and the simulation models, and then running NPE retrievals on each $x$. If the posteriors are well calibrated, then the parameter values $\theta$ that were used to generate the spectra $x$ should be contained in the $1-\alpha$ highest posterior density regions of the approximate posteriors $p_\phi(\theta | x)$  exactly $(1-\alpha)\%$ of the time. 
If the coverage probability is smaller than the credibility level $1 - \alpha$, then this indicates that the $1-\alpha$ highest posterior density regions are smaller than they should be, which is the sign of overconfident and usually unreliable posterior approximations. On the other hand, if the coverage probability is larger than the credibility level $1 - \alpha$, then this indicates that the highest posterior density regions are wider than they should be. In this case, the posterior approximations are said to be conservative. We argue that posterior approximations should rather be conservative to guarantee reliable and meaningful inferences, even when the approximations are not faithful. Indeed, wrongly excluding plausible parameter values of exoplanet spectra could lead to wrong conclusions about the actual nature of the exoplanet, while failing to exclude actually implausible parameter values would only result in a loss of statistical power. 
Figure~\ref{fig:coverage_plot} summarizes the expected coverage of $p_\phi(\theta|x)$ for credibility levels from $0$ to $1$. The coverage curve closely fits the diagonal, which indicates that the posterior distributions produced by NPE are well calibrated -- even though we note a trend for the posteriors to be very slightly underdispersed.

Unfortunately, running the same coverage diagnostic for a sampling algorithm such as MCMC or nested sampling is not possible within a reasonable computation time, since it requires the repeated construction of posterior distributions over many distinct random realizations $x$ in order to approximate the expectation in Eq.~\ref{eq:coverage}. For this reason, we cannot conclude whether \texttt{MultiNest} is computationally faithful in terms of expected coverage.  However, given that the approximate posterior distribution produced by \texttt{MultiNest} in Fig.~\ref{fig:results} is slightly narrower than the NPE posterior distribution, it suggests that \texttt{MultiNest} is slightly more underdispersed than NPE. This conclusion is in line with the analysis of \texttt{MultiNest} posterior distributions in \citet{martinez2022convolutional}, where 4000 retrievals were performed on simulated observations using CNNs and MultiNest, which on comparison suggest that MultiNest tends to underestimate the uncertainties of the parameter it retrieves.


\subsection{Computational cost}
\label{sec:exp-speedup}

One of the main advantages of neural posterior estimation is its amortization of the inference procedure. Once trained, inference does not require on-the-fly simulations and can be repeated several times with different observations at very low computational cost. We demonstrate the true potential of NPE by performing 1000 retrievals and comparing how long it would take for \texttt{MultiNest} to produce comparable results. The 1000 observations are produced by randomly sampling parameters values $\theta$ from the prior distribution and rendering them through the forward simulation model to produce $x \sim p(x|\theta)$. We then retrieve their corresponding approximate posterior distributions $p_\phi(\theta|x)$. A single retrieval consists of sampling \num{30740} posterior parameter vectors (as many as \texttt{MultiNest} yields) and rendering the corner plot, taking respectively 6 and 10 seconds in average. In total, 1000 retrievals would take approximately 4.5~hours. With the upfront generation of the dataset (17 hours on 1000 CPUs) and the training of the neural network (13 hours on a standard NVIDIA GTX 1080 Ti GPU), we reach a total computing time of 34.5~hours. By contrast, each retrieval with \texttt{MultiNest} takes around 134 hours on a cluster of 440 CPUs (totalling about \num{60000} CPU hours) so that retrieving atmospheric parameters on 1000 spectra would require an extrapolated time of \num{134000} hours (15 years). In summary, NPE is around \num{4000} times faster for a thousand retrievals, and almost \num{30000} times faster if we do not take the upfront generation and training into account.

It is important to note that the computational speedup comes with the overhead cost of building the training set (one per atmospheric model) and training NPE on it.
In our case, simulating a single parameter-spectrum pair $(\theta, f(\theta))$ takes around  5 seconds, which results in a total of \num{17000} CPU hours for the generation of the 12 millions pairs used in this study.
The actual wall-clock time, however, can be largely reduced by simulating the pairs in parallel on a large computing cluster, contrary to the on-the-fly and sequential simulations required in MCMC or nested sampling methods.  
In our case, the training set was generated in less than 17 hours using a cluster of 1000 CPUs. 
Generating as many samples may not be necessary in all cases, since sufficiently good performance is likely to be possible from smaller training sets.
Instead, the main challenge with amortized inference will be to identify a simulation model that is general enough to be applicable and valid in many situations, so that the whole training process does not need to be repeated for each individual case. This may be possible for studies focusing on specific classes of planets, such as hot Jupiters observed in transit, or self-luminous giant planets observed with direct imaging. We also note that, when performing retrievals on a single object, a given training data set can potentially be reused several times when exploring various levels of wavelength binning or different noise models in the retrieval. In this case, only the cost of the NPE training needs to be paid several times.

\section{Related work} 
\label{sec:related}

The closest work to our study is the recent work of \citet{yip2022sample}, which investigates variational inference and normalizing flows for atmospheric exoplanet retrieval.
In contrast to NPE, their approach is non-amortized, and variational inference is targeted at a single spectrum. 
For this reason, the expected KL divergence trick we use in Eq.~\ref{eq:loss} to bypass the sampling of the unknown posterior $p(\theta|x)$ is no longer applicable. Instead, the parameters of the normalizing flow are trained by maximizing a variational lower bound on the evidence $p(x)$, provided that the likelihood function associated with the simulation model is both tractable and differentiable. In NPE, none of these requirements are necessary -- the inference algorithm can be applied to any kind of simulation model, tractable or not, differentiable or not. 
Nevertheless, the amortization in NPE comes at the price of the upfront simulation of a large training set, whereas direct variational inference as in \citep{yip2022sample} only requires a limited number of on-the-fly simulations. 
These on-the-fly simulations, however, make inference significantly slower than the quasi-instantaneous inference produced by an already-trained normalizing flow. In particular, this prevents posterior diagnostics (as in Section~\ref{sec:exp-diagnostics}), as they become computationally very expensive.

Beyond exoplanet retrieval, NPE is used increasingly for inference problems found in astronomy. 
Close to exoplanet atmospheric retrieval, \citet{baso2022bayesian} and \citet{ramos2021approximate} use NPE to determine the thermodynamic and magnetic properties of solar and stellar atmospheres as well as their high-dimensional temperature maps. They similarly advocate for amortized and rapid parameter estimation if complex models are used to analyze the large amounts of data that the next generation of telescopes will produce. 
In gravitational wave science, \citet{dax2021real} use NPE for fast and accurate inference of the properties of binary black holes from gravitational waves. The inference time is reduced from 1 day using MCMC to 20s using NPE, making a strong case for inference in real-time. 
Similarly, \citet{zhang2020automating} and \citet{hahnaccelerated} use NPE to perform inference on binary microlensing events. Complex high-dimensional physical models result in time-consuming forward simulations and complex likelihood surfaces that MCMC methods find challenging to sample from. NPE offers a way to infer from an upcoming catalogue of binary events more accurately and in real-time. 
In astroparticle physics, \citet{PhysRevD.105.063017} use NPE to improve the characterization of the sources that contribute to the Fermi $\gamma$-ray Galactic Centre Excess (GCE), by directly sampling from high-dimensional $\gamma$-ray maps instead of defining a simplified and tractable likelihood function that loses some information. 
Likewise, \citet{bister2022inference} study the inference of cosmic-ray source properties from cosmic-ray observations on Earth. They conclude that inference with NPE provides accurate, fast, and verifiable results for a large phase space of the source parameters. 
Finally, as a last example, {\citet{kodi2020dynamical}} use NPE to characterize the dynamical mass of galaxy clusters directly from their 2d phase-space distributions. 

For the same reasons of efficiency, scalability, and testability, simulation-based inference algorithms beyond NPE are increasingly used across astronomy and other fields of science. 
Prominent algorithms include neural ratio estimation \citep{hermans2020likelihood,durkan2020contrastive}, which builds a surrogate of the likelihood-to-evidence ratio, and neural likelihood estimation \citep{papamakarios2016fast, alsing2018massive,papamakarios2019sequential}, which learns a fast and tractable surrogate of the likelihood.
    
\section{Conclusion} 
\label{sec:conclusion}

In this paper, we implemented a simulation-based inference algorithm called NPE to perform Bayesian retrievals of exoplanet atmospheres. Unlike the commonly used nested sampling and MCMC methods, which perform sequential sampling to construct a joint posterior of all model parameters using an explicit and tractable likelihood function, NPE relies on normalizing flows to estimate the posterior in an amortized way, without requiring an explicit or tractable likelihood. This offers several benefits over standard algorithms. 

First, NPE is time efficient due to amortization. The inference network needs to be trained only once, and the same network can be used to perform quasi-instantaneous retrievals over several observations without starting from scratch. We demonstrate this by performing $1000$ retrievals with synthetic observations constructed by sampling randomly from the prior. This procedure takes $34.5$ hours in total, leading to a computational speed up of $4000$ over \texttt{MultiNest}. The initial overhead cost of simulations is around $17$ hours, which is easily compensated as the number of observations increases. In the case where several simulation models $f$ need to be tested for the retrieval on a given observation, NPE still provides a speed up of over a factor 4 ($134/30$).

Second, NPE is scalable. Since the inference network is trained on the parameters of interest only, performance does not deteriorate as quickly as sampling-based algorithms that must navigate the full joint posterior over both the parameters of interest and the nuisance parameters. 
This is especially important for future simulation models that are likely to include a large number of nuisance parameters.

Lastly, NPE is testable. Since the inference of many observations takes only seconds to perform, one can easily check the validity of NPE by performing posterior predictive checks and producing coverage plots, which is almost impossible to achieve in the case of sequential algorithms. The results presented in this study confirm that NPE provides computationally faithful posteriors, without any simplifying assumption on the shape of the posterior, yet with a possible sign of being slightly underdispersed. While such tests cannot be performed with \texttt{Multinest} to provide a fair comparison, our mock retrievals suggest that NPE may be less underdispersed and more faithful than \texttt{Multinest}.


NPE’s computational speed up opens the possibility of efficiently retrieving atmospheric parameters from large datasets of exoplanet spectra. The speed up provided in the retrieval of individual spectra also enables the exploration of several different simulation models over limited observations in a reasonable time. The prospect of subjecting these retrievals to evaluation metrics like posterior predictive checks and coverage plots ensures a statistical rigour to the associated results. This establishes NPE as a robust algorithm to perform time-efficient retrievals in the future.

\begin{acknowledgements}
    This project has received funding from the European Research Council (ERC) under the European Union’s Horizon 2020 research and innovation programme (grant agreements No 819155 and 832428), and from the Wallonia-Brussels Federation (grant for Concerted Research Actions). 
    G.L. is recipient of the ULiège - NRB Chair on Big Data and is thankful for the support of the NRB.
\end{acknowledgements}

%
  \bibliographystyle{aa} 
  \bibliography{ref.bib} 
%
\onecolumn
\begin{appendix} 

\section{Normalizing flows 101}\label{normflow} 

Normalizing flows \citep{papamakarios2021normalizing} are invertible mappings transforming a simple probability distribution to a complex one. The change in the probability density of a random variable $u$ due to a invertible transformation $g$ is given by the change of variables theorem as 
$$ \log p(v) = \log p(u) - \log \left| \text{det} \frac{\partial g(u)}{\partial u} \right|, $$
where $v = g(u)$, and the determinant accounts for the change in volume between the two distributions. Because the transformation is invertible, the opposite direction $u = t(v)$ where $t = g^{-1}$ is also tractable. In this direction, the density ``flows'' from a complex distribution to a Normal distribution, hence the name ``normalizing flows'' (NFs).

To increase the expressiveness of NFs, parametric transformations can be stacked up as $v = g_{n} \circ g_{n-1} \circ \ldots \circ g_1 (u), $ which results in the probability density
$$\log p(v) = \log p(u) - \sum_{i=1}^{n} \log \left| \text{det} \frac{\partial g_i(u_{i-1})}{\partial u_{i-1}} \right| , $$
where $u_i = g_i (u_{i-1})$ and $u_0 = u$. 

\begin{figure*}[!h]
    \centering
    \includegraphics{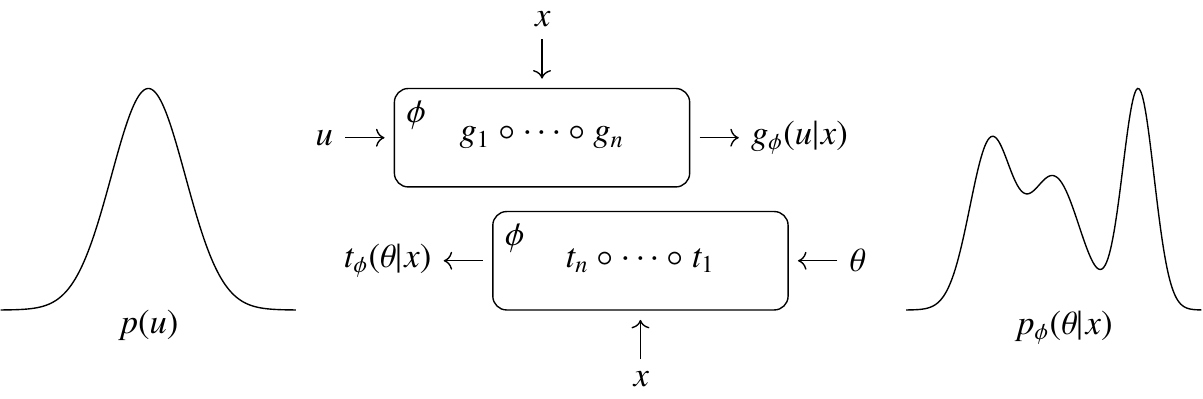}
    \caption{Transformation of a random variable $z$ with probability density $p(z)$ towards a variable $\theta$ with probability density $p_\phi(\theta|x)$. }
    \label{fig:nf}
\end{figure*}

In this work, we similarly model the posterior density $p_{\phi}(\theta|x)$ of the variable $\theta$ through a series of transformations of a Normal random variable $u$ with probability density $p(u)$, as illustrated in Fig.~\ref{fig:nf}. In the case of neural autoregressive flows, the transformations are invertible neural networks conditioned to the observation $x$.

\end{appendix}

\end{document}